\documentclass[prd,twocolumn,preprintnumbers,nofootinbib,tightenlines,superscriptaddress]{revtex4}

\usepackage{lipsum}

\pdfoutput=1
\usepackage[english]{babel}
\usepackage{amsmath,amssymb,amsfonts,bm,bbm,slashed,subdepth}
\usepackage{graphicx}
\usepackage[sort&compress]{natbib}
\usepackage{xcolor}
\usepackage[normalem]{ulem}
\usepackage{hyperref}
\usepackage{cleveref}
\definecolor{red}{rgb}{1.0, 0, 0}
\usepackage{enumerate}
\usepackage{epsfig, subfigure}
\usepackage{booktabs, tabularx}
\usepackage{units}
\usepackage{placeins}
\usepackage{multirow}
\usepackage{mathtools}
\usepackage{feynmp-auto}
\usepackage{slashed}
\usepackage{float}
\usepackage{setspace}

\usepackage{titlesec}
\titlespacing*{\section}{0pt}{15pt}{15pt}

\newcommand{\RNum}[1]{\uppercase\expandafter{\romannumeral #1\relax}}

\definecolor{myred}{rgb}{0.6078431372549019,0.11372549019607843,0.12549019607843137}
\definecolor{myblue1}{rgb}{0.16791, 0., 0.301671}
\definecolor{myblue2}{rgb}{0.28235299999999997, 0.1497248, 0.6790623333333333}
\definecolor{myblue3}{rgb}{0.26669366666666666, 0.550462, 0.926485}
\definecolor{myblue4}{rgb}{0.6711543333333333, 0.814616, 0.9359733333333333}
\definecolor{mypurp}{rgb}{0.761959, 0.470832, 0.940597}

\usepackage{bm,epstopdf,natbib,hyperref,color,verbatim,multirow,bm,tikz,xcolor}
\hypersetup{colorlinks=true,urlcolor=blue,citecolor=blue,linkcolor=blue,menucolor=blue,anchorcolor=blue,filecolor=blue}
\everymath{\displaystyle}
\definecolor{lime}{HTML}{A6CE39}
\DeclareRobustCommand{\orcidicon}{
	\begin{tikzpicture}
	\draw[lime, fill=lime] (0,0) 
	circle [radius=0.16] 
	node[white] {{\fontfamily{qag}\selectfont \tiny ID}};
	\draw[white, fill=white] (-0.0625,0.095) 
	circle [radius=0.007];
	\end{tikzpicture}
	\hspace{-3mm}
}
\foreach \x in {A, ..., Z}{\expandafter\xdef\csname orcid\x\endcsname{\noexpand\href{https://orcid.org/\csname orcidauthor\x\endcsname}
			{\noexpand\orcidicon}}
}

\begin{document}

\preprint{LA-UR-25-22499}

\title{Naturally Resonant Dark Matter from Extra Dimensions}

\author{Taegyu Lee\hspace{-1mm}\orcidA{}}
\email{taeglee@iu.edu}
\affiliation{Department of Physics, Indiana University, Bloomington, Indiana 47405, USA}

\author{Yu-Dai Tsai\hspace{-1mm}\orcidB{}}
\email{yudaitsai.academic@gmail.com}
\affiliation{The University of Manchester, Manchester M13 9PL, United Kingdom}
\affiliation{Los Alamos National Laboratory (LANL), Los Alamos, New Mexico 87545, USA}
\affiliation{School of Mathematics and Physical Sciences, University of Sheffield, Sheffield S3 7RH, United Kingdom}

\setlength{\unitlength}{0.7cm}
\begin{abstract} 
We explore the mass-resonance structure that naturally arises in extradimensional models. The resonance can enhance both dark matter annihilation and self-interactions. 
We demonstrate such a resonance structure by considering the fermionic dark matter and dark photon models on an $S^1/(Z_2 \times Z_2')$ orbifold. 
We also note that this model embeds a dark matter axial-vector coupling to the dark photon, thereby opening up the viable dark matter parameter space.
We then present the unique predictions for direct-detection experiments and accelerator searches.
\end{abstract}
\maketitle

\textbf{\emph{Introduction}} -- Despite overwhelming observational evidence for dark matter (DM) and its profound impact on particle physics, astrophysics, and cosmology, the detailed properties of DM remain unknown. More specifically, its interactions beyond gravity are yet to be discovered. One of the most compelling DM candidates, which features a predictive thermal history and interaction strength with Standard Model (SM) particles, is thermal DM.
Currently, direct-detection and accelerator experiments have explored a large portion of the parameter space of thermal DM models mediated through dark photons~\cite{Antel:2023hkf,Lee:2026djo}, 
except in scenarios where an enhancement of annihilation occurs due to the presence of a Breit-Wigner resonance~\cite{Feldman:2008xs,Ibe:2008ye,Feng:2017drg}. 
These phenomenological mass-resonance models need more underlying UV theories and theoretical motivations.

Resonant structures are ubiquitous in the Standard Model, and it has been proposed that DM could exhibit a similar resonant behavior~\cite{Tsai:2020vpi}. Such resonances can have intriguing implications, including opening up a valid parameter space by introducing resonance between mediators and DM~\cite{Ibe:2008ye,Feng:2017drg}. Furthermore, they can address small-scale structure problems by introducing self-interactions and resonances among DM particles~\cite{Chu:2018fzy}, which can also affect the self-freeze-out process~\cite{Kuflik:2015isi,Kuflik:2017iqs}. While such structures can arise from strongly interacting sectors or supersymmetry~\cite{Csaki:2022xmu}, the possibility that resonant DM candidates originate from extradimensional theories remains underexplored.

Extradimensional theories were first introduced to unify gravity and electromagnetism~\cite{Appelquist:1987nr}, and the development of string theory later provided a consistent framework for quantum gravity, wherein extra dimensions are compactified at scales near the Planck length. Furthermore, to address the hierarchy problem and the observed weakness of gravity, models with flat extra dimensions were proposed~\cite{Arkani-Hamed:1998jmv}, followed by warped extradimensional scenarios~\cite{Randall:1999ee}. Unlike some of these frameworks, where only gravity propagates in the extra dimensions, the universal extra dimension (UED) model allows all SM particles to traverse the extradimensional space~\cite{Appelquist:2000nn,Datta:2010us}. 
In this broad landscape, the implications of extra dimensions for resonant structures, particularly in connection with DM, remain an area of ongoing investigation.

This work represents one of the first attempts to investigate resonant structures within extradimensional theories and their application to DM. Specifically, we explore the resonance structure of DM in the framework of simple Kaluza-Klein (KK) models that naturally accommodate resonant behavior. Our model is inspired by the five-dimensional (5D) UED framework, where the DM and dark photon KK modes are embedded in an extra dimension compactified on an $S^1/Z_2$ or $S^1/(Z_2 \times Z_2')$ orbifold. 
Unlike UED~\cite{Servant:2002aq,Bertone:2002ms,Servant:2002hb,Servant:2014lqa}, we will focus on the scenarios in which only the dark sector propagates in the extra dimension while the SM is localized on the 4D spacetime.
We demonstrate that this model can naturally achieve the desired resonance level between DM and dark photons with minimal fine-tuning, thereby opening up a new region of parameter space for the kinetic mixing parameter of invisible dark photons.
Additionally, we briefly discuss the parameter space required to achieve resonance levels capable of mitigating small-scale structure problems.

\textbf{\emph{Fermionic Dark Matter Model}} -- Throughout this paper, we adopt the convention that the coordinates in (3+1) dimensions are denoted by $x^\mu$ (where $\mu = 0, 1, 2, 3$), while $y$ represents the fifth extra dimension. The metric signature used is mostly negative, $g_{MN} = (+----)$. 

We consider a model that is characterized by the size of the orbifold, $R$, with a compactification scale $M_c = 1/R$. This model consists of a fermion field in addition to the dark photon gauge field, where these KK modes are compactified on an $S^1/(Z_2\times Z'_2)$ orbifold. The matter content of the model consists of an $SU(2)_L$ singlet fermion, $E(x,y)$, and a dark photon ($\gamma_D$), $B_N(x,y)$, where $N = \mu, 5$. Each field obeys specific boundary condition on each boundary point. A summary of the boundary conditions is presented in Table~\ref{table:BCS1Z2Z2}. The Supplemental Material provides further details on $S^1/(Z_2\times Z'_2)$ and compares the convention used in this paper with those in the literature, including Ref.~\cite{Barbieri:2000vh}.
\begin{table}[h!]
\centering
\begin{tabular}{|c | c | c | c | } 
 \hline
 States & $y = 0$ & $y = \pi R/2$ & $y = \pi R$ \\ [1ex]
 \hline
 RH Singlet Fermion & Neumann & Dirichlet & Neumann \\
 \hline
 LH Singlet Fermion & Dirichlet & Neumann & Dirichlet \\
 \hline
 Dark Photon ($B_\mu$) & Neumann & Neumann & Neumann \\ 
 \hline
 Dark Photon ($B_5$) & Dirichlet & Dirichlet & Dirichlet \\ 
 \hline
\end{tabular}
\caption{Summary of the boundary conditions imposed on KK particles.}
\label{table:BCS1Z2Z2}
\end{table}
The implications of the conditions are noteworthy. To satisfy these boundary conditions, dark photons must consist of only $n$-even modes, while right-handed (RH) singlet fermions and left-handed (LH) singlet fermions must consist of only $n$-odd modes. Furthermore, we imply the zero-flux condition~\cite{Kobayashi:2022xsk}, $\oint B_{\mu} dy=0$, to eliminate the zeroth mode of the gauge field $B^{(0)}_{\mu}$ as follows: 
\begin{align}
    &E(x,y) = \sqrt{\frac{4}{\pi R}}\Bigg[ \Bigg.\sum^\infty_{n=1} E^{(2n-1)}_R(x)\cos{\left(\frac{(2n-1)y}{R}\right)} \nonumber \\ & \;\;\;\;\;\;\;\;\;\;\;\;\;\;\;\;\;\; + E^{(2n-1)}_L(x)\sin{\left(\frac{(2n-1)y}{R}\right)} \vphantom{\sum^\infty_{n=1}}\Bigg. \Bigg], \\
    &B_{\mu}(x,y) = \sqrt{\frac{4}{\pi R}}\left[\sum^{\infty}_{n=1}B^{(2n)}_\mu(x)\cos\left(\frac{2ny}{R}\right)\right], \nonumber \\
    &B_5(x,y) = \sqrt{\frac{4}{\pi R}}\sum^{\infty}_{n=1}B^{(2n)}_5(x)\sin\left(\frac{2ny}{R}\right).
\end{align}
The superscript $n$ represents the KK number. 

The Lagrangian in (3+1) dimensions is constructed by integrating out the extradimensional coordinate $y$. The Lagrangians involving fermions and dark photons are 
\begin{equation}
    \mathcal{L}_E = \int^{\pi R/2}_0 dy \left[i\bar{E}\Gamma^M D_M E\right],
\label{eqn:Lag_model12}
\end{equation}
\begin{equation}
     \mathcal{L}_{\gamma_D} = \int^{\pi R/2}_0 dy \left(-\frac{1}{4}B_{MN}B^{MN}-\frac{1}{2}m_B^2B_NB^N\right),
\label{eqn:Lag_model1}
\end{equation}
where $B_{MN}$ is the 5D field strength tensor of the dark photon. The covariant derivative acting on the field in the Lagrangian is given by 
\begin{eqnarray}
    D_M E &=& \left(\partial_M + i Q_E g_1^{(5)} B_M\right) E,
\end{eqnarray}
where $B_M$ is the KK dark photon, and $Q_E$ and $g^{(5)}_1$ are the associated charge and 5D gauge coupling, respectively. Throughout this work, $Q_E$ is set to unity. Here, $g^{(5)}_1$ is a dimensionful quantity in 5D and is related to the dimensionless 4D gauge coupling, $g_1$, by
\begin{equation}
    g_1 = \frac{g_1^{(5)}}{\sqrt{L}},
\label{4D_gauge}
\end{equation}
where $L = \pi R/2$ for the $S^1/(Z_2\times Z_2')$ orbifold. The $\gamma$ matrices and partial derivative operators in 5D are given by,
\begin{eqnarray}
    \Gamma^M &=& (\gamma^\mu, i\gamma^5), \\
    \partial_M &=& (\partial_\mu, \partial_y).
\end{eqnarray}

The fifth component of the partial derivative, $\partial_y$, introduces an additional contribution to the mass for each KK mode, in addition to the bare mass. Therefore, the total mass of the $n$th mode for each particle is
\begin{eqnarray}
    m^{(n)}_E &=& \frac{n}{R}, \\
    m^{(n)}_{\gamma_D} &=& \sqrt{m_B^2 + \frac{n^2}{R^2}},
\label{mass_tree}
\end{eqnarray}
at tree level. Here, we assume that a bare mass term for the fermion is forbidden due to the lack of a Higgs mechanism, while the dark photon acquires a bare mass through a different mechanism. The three-point vertex essential for the resonant structure, $EE \rightarrow \gamma_D$, is given in Eq.~\eqref{eqn:Lag_model12}. Due to the structure of the model, chirality-conserving vertices occur only with $B_\mu$, while chirality-flipping vertices occur only with $B_5$. 

It can be shown that the only stable particle in the model is $E^{(1)}$. Given that $B^{(2)}$ decays into two electrons, the decay channels for the remaining particles are as follows: $E^{(2n+1)}\rightarrow E^{(2n-1)} B^{(2)*}\rightarrow E^{(2n-1)}e^+e^-$ and $B^{(2n+2)}\rightarrow E^{(2n-1)} E^{(3)*} \rightarrow E^{(2n-1)} E^{(1)} B^{(2)*} \rightarrow E^{(2n-1)} E^{(1)} e^+ e^-$, where the superscript “*” denotes an off-shell particle. 

\begin{figure}
  \includegraphics[scale=1]{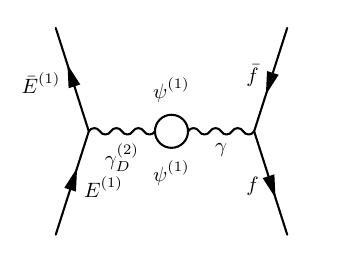}
\caption{Dark matter interaction with SM.
}
\label{Fig:SMresonance}
\end{figure}

\textbf{\emph{Resonant structures and interactions with SM}} -- Here, we briefly discuss the sub-GeV thermal DM, in our model, that freezes out to produce the correct DM relic abundance. The interaction with the SM can be achieved through kinetic mixing between dark photons and SM photons. This kinetic mixing arises through particles charged under both $U(1)_{\rm em}$ and $U(1)_D$. For instance, if $\psi$ is a fermion charged under both $U(1)_{\rm em}$ and $U(1)_D$, and behaves as a KK particle similar to $E$, then a diagram (shown in Fig.~\ref{Fig:SMresonance})
enables  
the interaction of DM with the SM. A key feature of this diagram is the resonance effect during DM annihilation into SM fermions.
To examine the resonance structure in detail, we must analyze the radiative corrections to the masses, as these can lead to substantial changes in the mass spectrum.

\begin{figure}
  \includegraphics[scale=1]{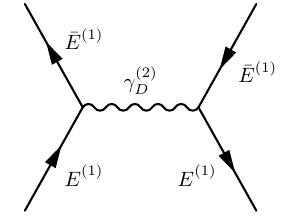}
\caption{A diagram that can lead to resonant structure in the self-interaction.}
\label{Fig:Selfresonance}
\end{figure}

\textbf{\emph{Potential resonant self-interactions}} --
Studies have shown that small-scale structure problems could be mitigated by introducing self-interactions among DM particles~\cite{Kaplinghat:2015aga}. Additionally, it has been proposed that resonant self-interacting DM could help reconcile simulation results with experimental data, thereby reducing existing tensions~\cite{Chu:2018fzy,Tsai:2020vpi}. Achieving this, however, requires a strong resonance effect, which may necessitate some fine-tuning of parameters. The desired resonance level, $\epsilon_R$, specified in Eq.~(\ref{eps_R}), should range between $10^{-8}$ and $10^{-4}$. This level of resonance can be achieved, with minimal fine-tuning, when the 4D gauge coupling $g_1$ is less than $0.01$. A brief discussion on small 4D gauge coupling is provided at the end of the next section.
 The relevant resonant structure is illustrated in the diagram in Fig.~\ref{Fig:Selfresonance}.

\textbf{\emph{Radiative Corrections to Masses and the Level of Resonance}} -- In this section, we discuss the one-loop radiative mass corrections to the states of interest, $B^{(2)}$ and $E^{(1)}$, and examine how these corrections impact the resonance structure.
General prescriptions for calculating the radiative corrections in extradimensional models are provided in Refs.~\cite{Freitas:2017afm,Georgi:2000ks,Carena:2002me,Cheng:2002iz}, and we focus on our special consideration below. The mass corrections to the states of interest at one loop level are given by
\begin{eqnarray}
    m_{B^{(2)}}' &=& \sqrt{m_B^2+\left(\frac{2}{R}\right)^2-\frac{g_1^2\zeta(3)}{16\pi^4R^2}}, \\
    m_{E^{(1)}}' &=& \frac{1}{R}\left(1+\frac{9}{64\pi^2}g_1^2\ln{\frac{\Lambda^2}{\mu^2}}\right).
\end{eqnarray}

We define the ``level of resonance", $\epsilon_R$~\cite{Feng:2017drg}, as follows:
\begin{equation}
    \epsilon_R \equiv \frac{\left(m_{B^{(2)}}'\right)^2-\left(2m_{E^{(1)}}'\right)^2}{\left(2m_{E^{(1)}}'\right)^2}.
\label{eps_R}
\end{equation}

A positive value of $\epsilon_R$ ($\epsilon_R > 0$) corresponds to the case of an invisible dark photon, while a negative value ($\epsilon_R < 0$) corresponds to a visible dark photon. 
In the Supplemental Material, we discuss and plot the level of resonance as a function of $m_B$ for fixed $\Lambda / \mu$, with other parameters set to $g_1 = 0.1,1$ and $1 / R = 1~\text{GeV}$. It is particularly noteworthy that, for $g_1=0.1$, when $m_B \sim 1 / R$, the resonance level remains $\epsilon_R \lesssim \mathcal{O}(0.1)$ across a wide range of cutoff scales, $\Lambda / \mu$. This feature enables resonance over a large parameter space while consistently preserving the invisibility of the dark photon. Compared with the SM $U(1)$ gauge coupling, $g' = 0.357$, both $g_1 = 0.1$ and $g_1 = 1$ represent viable benchmark choices.

Moreover, very small gauge couplings may be considered unnatural if the 4D gauge coupling is viewed as a fundamental parameter. However, in the context of 5D, $g_1^{(5)}$ is not a fundamental parameter since it is dimensionful. This implies that there is no reason to consider $g_1^{(5)} \sim \sqrt{L}$ as more natural (for a related discussion, see Ref.~\cite{Sakamura:2014aja}). Therefore, one can always select $g_1^{(5)}$ and $L$ such that the 4D gauge coupling, $g_1 = g_1^{(5)}/\sqrt{L}$, is very small.
This approach further enhances the level of resonance, potentially enabling the realization of resonant self-interacting DM to help alleviate small-scale structure problems~\cite{Chu:2018fzy}.

\textbf{\emph{Resonance Freeze-out of the Relic Abundance}} --
\begin{figure*}[ht]
\includegraphics[scale=0.5]{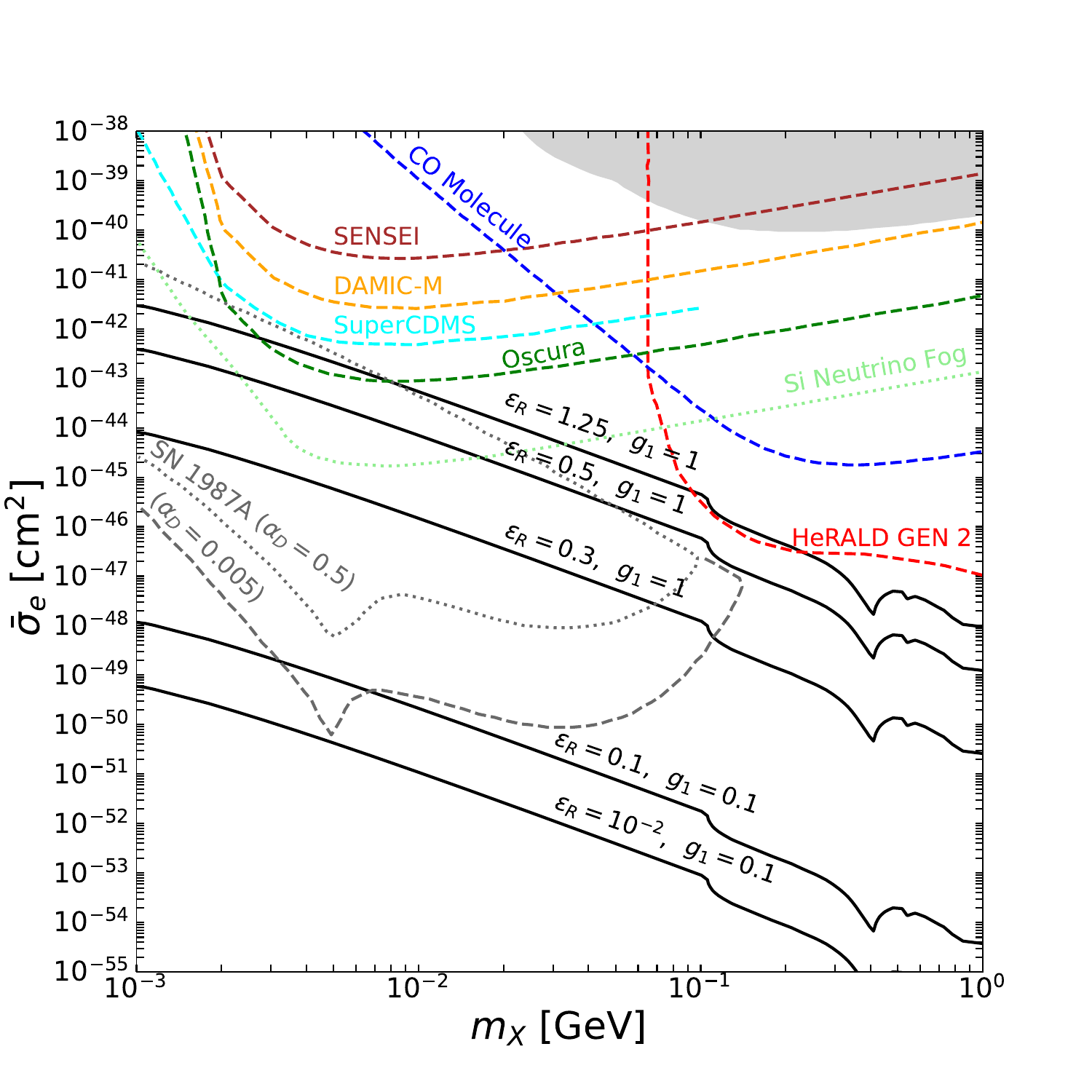}
\caption{
The solid black contours correspond to the projections of the model for different values of $\epsilon_R$, where $\epsilon_R = 1.25$ corresponds to the non-resonant case when $m_{B^{(2)}}' = 3 m_{E^{(1)}}'$. For $\epsilon_R=1.25,\;0.5,\;0.3$, we set $g_1=1\;(\alpha_D\sim0.04)$ and for $\epsilon_R=0.1,\;0.01$, we set $g_1=0.1\;(\alpha_D\sim0.0004)$.  The gray-shaded region is the excluded region. We also include projected sensitivities, dashed lines, from future experiments, including Sub-Electron-Noise Skipper CCD Experimental Instrument (SENSEI)~\cite{Essig:2015cda}, Dark Matter in CCDs at Modane (DAMIC-M)~\cite{Castello-Mor:2020jhd},  Oscura \cite{Oscura:2023qik}, super cryogenic dark matter search (SuperCDMS)-SNOLAB (SuperCDMS) \cite{SuperCDMS:2022kse},  DM scattering off of molecules (CO Molecule) \cite{Essig:2019kfe}, and a superfluid Helium experiment (HeRALD (helium roton apparatus for light dark matter))~\cite{Hertel:2018aal}. The dotted green curve corresponds to the neutrino fog for silicon detectors (Si Neutrino Fog)~\cite{Carew:2023qrj}. The dotted/dashed gray curve is the constraint from supernova SN 1987A for $\alpha_D=0.5/0.005$ ~\cite{Chang:2016ntp,Chang:2018rso}. For SuperCDMS, we present only the projection curve from the cited paper, but the sensitivity can be extended to higher masses.
}
\label{expmodel}
\end{figure*}
This section explores the DM parameter space where the mass resonance can be realized as a concrete model. The dark photon decays invisibly to the dark sector, meaning $m_{B^{(2)}}' > 2m_{E^{(1)}}'$. Therefore, it is sufficient to identify the parameter space where $0 < \epsilon_R < \delta$, where $\delta\lesssim 0.5$, to ensure a sufficient resonance effect.  

Here, we project our model to the experimental constraints for the case of invisible dark photons, following related thermal-target studies of vector-portal dark matter~\cite{Fitzpatrick:2020vba,Fitzpatrick:2021cij}.
The 4D Lagrangian that is responsible for the interaction given in Fig.~\ref{Fig:SMresonance} is given as 
\begin{eqnarray}
        \mathcal{L} &\supset& -g_1'\bar{E}^{(1)}\gamma^\mu \gamma^5 E^{(1)} B_\mu^{(2)} - \Sigma_{f} \kappa e q_f \bar{f}\gamma^\mu fB_{\mu}^{(2)} \nonumber \\
        && -g_1'\bar{E}^{(1)} \gamma^5 E^{(1)} B_5^{(2)},
\end{eqnarray}
where $g_1'=g_1/\sqrt{2}$, $\kappa$, $e$, and $q_f$ are 4D gauge coupling, kinetic mixing parameter, electric charge, and $U(1)_{\rm em}$ quantum number, respectively. The dark fine structure constant, $\alpha_D$, is constructed from $g_1'$.
Interestingly, the coupling between DM and dark photon is given by axial vector coupling. This kind of coupling is general in the UED model where the first mode of fermion couples to the second mode of vector boson via axial vector coupling, see Ref.~\cite{Datta:2010us}. This leads to interesting phenomenology that we highlight at the end of this section. 
Note that the third term does not contribute to the annihilation cross section since the fifth component of the second mode of dark photon cannot mix with the SM photon. This is because there is no two-point vertex available between the scalar field ($B_5$) and the vector field (SM photon). 

We show the detailed calculation of the relic abundance in the supplemental material.

\textbf{\emph{Targets for Direct-Detection Experiments}} --
The DM-dark photon coupling will induce the DM-electron scattering cross section that can be tested by future direct detection measurements. We present our model predictions in Fig.~\ref{expmodel} using the reference cross section, $\bar{\sigma}_e$, that has been defined in Refs.~\cite{Essig:2011nj,Essig:2015cda},
\begin{equation}
    \bar{\sigma}_e = \frac{16\pi \mu_{e,E^{(1)}}^2 \alpha\alpha_D \kappa^2v^2}{(( m_{B^{(2)}}')^2+\alpha^2m_e^2)^2},
\end{equation}
where $\mu_{e,E^{(1)}}$ is the reduced mass of electron and dark matter and $v=10^{-3}$ is the typical velocity of DM halo. Interestingly, we have additional velocity suppression in the cross section. This is because the interaction between the dark photon and the DM, in this model, is given by axial vector coupling, while the effective interaction between the dark photon and the SM fermions remains as a vector coupling.

In Fig.~\ref{expmodel}, we present the naturally resonant dark matter predictions with different levels of resonances, along with limits and projections from direct-detection experiments. The solid black contours shown in Fig.~3 correspond to parameter points that reproduce the observed total dark matter abundance. For a dark photon mass of $m'_{B^{(2)}} = 100~\mathrm{MeV}$, which is related to $m_X$ through Eq.~(\ref{eps_R}), we find $2\times 10^{-13} < \kappa^2 < 3\times 10^{-7}$ over the range of $\epsilon_R$ values considered. Smaller values of $\kappa$ correspond to smaller values of $\epsilon_R$. The existing limits are from experiments including DAMIC-M \cite{DAMIC-M:2023gxo}, DAMIC-SNOLAB (DAMIC at SNOLAB, i.e., Dark Matter in CCDs at SNOLAB)~\cite{DAMIC:2019dcn}, SENSEI \cite{SENSEI:2020dpa}, Exp\'erience pour DEtecter Les Wimps En Site Souterrain \cite{EDELWEISS:2020fxc}, SuperCDMS \cite{SuperCDMS:2018mne}, CDEX-10 (China dark matter experiment) \cite{CDEX:2022kcd}, DarkSide-50 \cite{DarkSide:2022knj}, XENON1T \cite{XENON:2019gfn,XENON:2021qze}, PandaX-II~\cite{PandaX-II:2021nsg}, and and from the energy loss of supernova SN 1987A~\cite{Chang:2016ntp}. The supernova SN 1987A bound shown in Fig.~3 should be interpreted only as a reference. This is because our projections correspond to $\alpha_D = 0.04, 0.0004$, and a proper, consistent derivation of the corresponding supernova constraints for this setup would require a dedicated analysis beyond the scope of this work. Additionally, we show projected future sensitivities from SuperCDMS-SNOLAB \cite{SuperCDMS:2022kse}, SENSEI and DAMIC-M \cite{Essig:2015cda,Castello-Mor:2020jhd}, Oscura \cite{Oscura:2023qik}, DM scattering off of molecules \cite{Essig:2019kfe}, and HeRALD~\cite{Hertel:2018aal}, along with the neutrino fog specific for silicon detectors~\cite{Carew:2023qrj}. More specifically, our model, with a minimal level of resonance, may be discovered by the Oscura experiment or the second generation of the HeRALD experiment. Furthermore, higher resonance level scenarios could also be potentially probed by third- and fourth-generation HeRALD experiments. 
Accelerator searches such as Belle II~\cite{Belle-II:2018jsg}, the NA64 experiment at CERN (NA64)~\cite{Gninenko:2019qiv}, forward liquid argon experiment~\cite{Anchordoqui:2021ghd}, mini Beam Dump eXperiment~\cite{Battaglieri:2020lds}, light dark matter eXperiment~\cite{Akesson:2022vza,LDMX:2018cma}, and proposed millicharged/fractionally charged particle searches at intensity-frontier facilities~\cite{Bailloeul:2025fde}, and LongQuest-inspired inelastic-dark-matter searches~\cite{Tsai:2019buq} can potentially discover or constrain the DM model we presented, which is currently not constrained by limits from BABAR~\cite{BaBar:2017tiz}, Compact Muon Solenoid~\cite{CMS:2021far}, NA$64e$~\cite{Andreev:2021fzd}, E137~\cite{Batell:2014mga}, liquid scintillator neutrino detector~\cite{deNiverville:2011it}, and mini booster neutrino experiment~\cite{MiniBooNEDM:2018cxm}. 

\begin{acknowledgements}

The authors thank Hitoshi Murayama, Radovan Dermisek, Jonathan Feng, Keith Hermanek, Manoj Kaplinghat, and Maxim Perelstein for fruitful discussions. We are especially grateful to Geraldine Servant and Tim Tait for insightful discussions on universal extra dimensions and the radiative corrections to KK mode masses.

This research is partially supported by LANL's Laboratory Directed Research and Development (LDRD) program. YDT thanks the generous support from the LANL Director's Fellowship.
This research is partially supported by the U.S. National Science Foundation (NSF) Theoretical Physics Program, Grant No.~PHY-1915005. This research was supported in part by grant NSF PHY-2309135 to the Kavli Institute for Theoretical Physics (KITP).
This work was partially performed at the Aspen Center for Physics, supported by National Science Foundation grant No.~PHY-2210452. This research was partly supported by the NSF under Grant No.~NSF PHY-1748958.  YDT is supported by the Dorothy Hodgkin Fellowship, funded by the Royal Society, United Kingdom, and appreciates the start-up support from the University of Manchester and University of Sheffield.
    
\end{acknowledgements}

\bibliography{final}{}

@article{Antel:2023hkf,
    author = "Antel, C. and others",
    title = "{Feebly-interacting particles: FIPs 2022 Workshop Report}",
    eprint = "2305.01715",
    archivePrefix = "arXiv",
    primaryClass = "hep-ph",
    reportNumber = "CERN-TH-2023-061, DESY-23-050, FERMILAB-PUB-23-149-PPD, INFN-23-14-LNF, JLAB-PHY-23-3789, LA-UR-23-21432, MITP-23-015",
    doi = "10.1140/epjc/s10052-023-12168-5",
    journal = "Eur. Phys. J. C",
    volume = "83",
    number = "12",
    pages = "1122",
    year = "2023"
}

@article{Fitzpatrick:2021cij,
    author = "Fitzpatrick, Patrick J. and Liu, Hongwan and Slatyer, Tracy R. and Tsai, Yu-Dai",
    title = "{New thermal relic targets for inelastic vector-portal dark matter}",
    eprint = "2105.05255",
    archivePrefix = "arXiv",
    primaryClass = "hep-ph",
    reportNumber = "MIT-CTP/5298, FERMILAB-PUB-21-189-AE-T",
    doi = "10.1103/PhysRevD.106.083507",
    journal = "Phys. Rev. D",
    volume = "106",
    number = "8",
    pages = "083507",
    year = "2022"
}

@article{Bailloeul:2025fde,
    author = "Bailloeul, Leo and Citron, Matthew and Cui, Yanou and Foroughi-Abari, Saeid and Hwang, Insung and Li, Fengyi and Tsai, Yu-Dai and Liu, Ming Xiong and Gunthoti, Kranti and Yoo, Jae Hyeok",
    title = "{Dedicated Searches for Millicharged Particles at Intensity-Frontier Facilities: SpinQuest and SHiP}",
    eprint = "2512.11027",
    archivePrefix = "arXiv",
    primaryClass = "hep-ph",
    month = "12",
    year = "2025"
}

@article{Tsai:2019buq,
    author = "Tsai, Yu-Dai and deNiverville, Patrick and Liu, Ming Xiong",
    title = "{Dark Photon and Muon $g-2$ Inspired Inelastic Dark Matter Models at the High-Energy Intensity Frontier}",
    eprint = "1908.07525",
    archivePrefix = "arXiv",
    primaryClass = "hep-ph",
    reportNumber = "FERMILAB-PUB-19-393-A-PPD",
    doi = "10.1103/PhysRevLett.126.181801",
    journal = "Phys. Rev. Lett.",
    volume = "126",
    number = "18",
    pages = "181801",
    year = "2021"
}

@article{Fitzpatrick:2020vba,
    author = "Fitzpatrick, Patrick J. and Liu, Hongwan and Slatyer, Tracy R. and Tsai, Yu-Dai",
    title = "{New pathways to the relic abundance of vector-portal dark matter}",
    eprint = "2011.01240",
    archivePrefix = "arXiv",
    primaryClass = "hep-ph",
    reportNumber = "FERMILAB-PUB-20-560-AE-PPD-T, MIT-CTP/5252",
    doi = "10.1103/PhysRevD.106.083517",
    journal = "Phys. Rev. D",
    volume = "106",
    number = "8",
    pages = "083517",
    year = "2022"
}

@article{Lee:2026djo,
    author = "Lee, Junseok and Takahashi, Fuminobu and Tsai, Yu-Dai",
    title = "{Number Theory in Quantum Physics: Minicharged Particles and the Prouhet-Tarry-Escott Problem}",
    eprint = "2603.12320",
    archivePrefix = "arXiv",
    primaryClass = "hep-ph",
    reportNumber = "TU-1300",
    month = "3",
    year = "2026"
}

@article{Feldman:2008xs,
    author = "Feldman, Daniel and Liu, Zuowei and Nath, Pran",
    title = "{PAMELA Positron Excess as a Signal from the Hidden Sector}",
    eprint = "0810.5762",
    archivePrefix = "arXiv",
    primaryClass = "hep-ph",
    doi = "10.1103/PhysRevD.79.063509",
    journal = "Phys. Rev. D",
    volume = "79",
    pages = "063509",
    year = "2009"
}

@article{Ibe:2008ye,
    author = "Ibe, Masahiro and Murayama, Hitoshi and Yanagida, T. T.",
    title = "{Breit-Wigner Enhancement of Dark Matter Annihilation}",
    eprint = "0812.0072",
    archivePrefix = "arXiv",
    primaryClass = "hep-ph",
    reportNumber = "SLAC-PUB-13479, IPMU08-0097",
    doi = "10.1103/PhysRevD.79.095009",
    journal = "Phys. Rev. D",
    volume = "79",
    pages = "095009",
    year = "2009"
}

@article{Feng:2017drg,
    author = "Feng, Jonathan L. and Smolinsky, Jordan",
    title = "{Impact of a resonance on thermal targets for invisible dark photon searches}",
    eprint = "1707.03835",
    archivePrefix = "arXiv",
    primaryClass = "hep-ph",
    reportNumber = "UCI-TR-2017-06",
    doi = "10.1103/PhysRevD.96.095022",
    journal = "Phys. Rev. D",
    volume = "96",
    number = "9",
    pages = "095022",
    year = "2017"
}

@article{Tsai:2020vpi,
    author = "Tsai, Yu-Dai and McGehee, Robert and Murayama, Hitoshi",
    title = "{Resonant Self-Interacting Dark Matter from Dark QCD}",
    eprint = "2008.08608",
    archivePrefix = "arXiv",
    primaryClass = "hep-ph",
    reportNumber = "FERMILAB-PUB-20-365-AE-T",
    doi = "10.1103/PhysRevLett.128.172001",
    journal = "Phys. Rev. Lett.",
    volume = "128",
    number = "17",
    pages = "172001",
    year = "2022"
}

@article{Chu:2018fzy,
    author = "Chu, Xiaoyong and Garcia-Cely, Camilo and Murayama, Hitoshi",
    title = "{Velocity Dependence from Resonant Self-Interacting Dark Matter}",
    eprint = "1810.04709",
    archivePrefix = "arXiv",
    primaryClass = "hep-ph",
    reportNumber = "DESY-18-176",
    doi = "10.1103/PhysRevLett.122.071103",
    journal = "Phys. Rev. Lett.",
    volume = "122",
    number = "7",
    pages = "071103",
    year = "2019"
}

@article{Kuflik:2015isi,
    author = "Kuflik, Eric and Perelstein, Maxim and Lorier, Nicolas Rey-Le and Tsai, Yu-Dai",
    title = "{Elastically Decoupling Dark Matter}",
    eprint = "1512.04545",
    archivePrefix = "arXiv",
    primaryClass = "hep-ph",
    doi = "10.1103/PhysRevLett.116.221302",
    journal = "Phys. Rev. Lett.",
    volume = "116",
    number = "22",
    pages = "221302",
    year = "2016"
}

@article{Kuflik:2017iqs,
    author = "Kuflik, Eric and Perelstein, Maxim and Lorier, Nicolas Rey-Le and Tsai, Yu-Dai",
    title = "{Phenomenology of ELDER Dark Matter}",
    eprint = "1706.05381",
    archivePrefix = "arXiv",
    primaryClass = "hep-ph",
    doi = "10.1007/JHEP08(2017)078",
    journal = "JHEP",
    volume = "08",
    pages = "078",
    year = "2017"
}

@article{Csaki:2022xmu,
    author = "Cs{\'a}ki, Csaba and Gomes, Andrew and Hochberg, Yonit and Kuflik, Eric and Langhoff, Kevin and Murayama, Hitoshi",
    title = "{Super-resonant dark matter}",
    eprint = "2208.07882",
    archivePrefix = "arXiv",
    primaryClass = "hep-ph",
    doi = "10.1007/JHEP11(2022)162",
    journal = "JHEP",
    volume = "11",
    pages = "162",
    year = "2022"
}

@book{Appelquist:1987nr,
    editor = "Appelquist, T. and Chodos, A. and Freund, P. G. O.",
    title = "{MODERN KALUZA-KLEIN THEORIES}",
    year = "1987"
}

@article{Arkani-Hamed:1998jmv,
    author = "Arkani-Hamed, Nima and Dimopoulos, Savas and Dvali, G. R.",
    title = "{The Hierarchy problem and new dimensions at a millimeter}",
    eprint = "hep-ph/9803315",
    archivePrefix = "arXiv",
    reportNumber = "SLAC-PUB-7769, SU-ITP-98-13",
    doi = "10.1016/S0370-2693(98)00466-3",
    journal = "Phys. Lett. B",
    volume = "429",
    pages = "263--272",
    year = "1998"
}

@article{Randall:1999ee,
    author = "Randall, Lisa and Sundrum, Raman",
    title = "{A Large mass hierarchy from a small extra dimension}",
    eprint = "hep-ph/9905221",
    archivePrefix = "arXiv",
    reportNumber = "MIT-CTP-2860, PUPT-1860, BUHEP-99-9",
    doi = "10.1103/PhysRevLett.83.3370",
    journal = "Phys. Rev. Lett.",
    volume = "83",
    pages = "3370--3373",
    year = "1999"
}

@article{Appelquist:2000nn,
    author = "Appelquist, Thomas and Cheng, Hsin-Chia and Dobrescu, Bogdan A.",
    title = "{Bounds on universal extra dimensions}",
    eprint = "hep-ph/0012100",
    archivePrefix = "arXiv",
    reportNumber = "YCTP-P12-00, EFI-2000-48, EFI--2000-48",
    doi = "10.1103/PhysRevD.64.035002",
    journal = "Phys. Rev. D",
    volume = "64",
    pages = "035002",
    year = "2001"
}

@article{Datta:2010us,
    author = "Datta, AseshKrishna and Kong, Kyoungchul and Matchev, Konstantin T.",
    title = "{Minimal Universal Extra Dimensions in CalcHEP/CompHEP}",
    eprint = "1002.4624",
    archivePrefix = "arXiv",
    primaryClass = "hep-ph",
    reportNumber = "HRI-P-10-02-001, RECAPP-HRI-2010-006",
    doi = "10.1088/1367-2630/12/7/075017",
    journal = "New J. Phys.",
    volume = "12",
    pages = "075017",
    year = "2010"
}

@article{Servant:2002aq,
    author = "Servant, Geraldine and Tait, Timothy M. P.",
    title = "{Is the lightest Kaluza-Klein particle a viable dark matter candidate?}",
    eprint = "hep-ph/0206071",
    archivePrefix = "arXiv",
    reportNumber = "ANL-HEP-PR-02-032, EFI-02-74",
    doi = "10.1016/S0550-3213(02)01012-X",
    journal = "Nucl. Phys. B",
    volume = "650",
    pages = "391--419",
    year = "2003"
}

@article{Bertone:2002ms,
    author = "Bertone, Gianfranco and Servant, Geraldine and Sigl, Guenter",
    title = "{Indirect detection of Kaluza-Klein dark matter}",
    eprint = "hep-ph/0211342",
    archivePrefix = "arXiv",
    reportNumber = "ANL-HEP-PR-02-099, EFI-02-52",
    doi = "10.1103/PhysRevD.68.044008",
    journal = "Phys. Rev. D",
    volume = "68",
    pages = "044008",
    year = "2003"
}

@article{Servant:2002hb,
    author = "Servant, Geraldine and Tait, Timothy M. P.",
    title = "{Elastic Scattering and Direct Detection of Kaluza-Klein Dark Matter}",
    eprint = "hep-ph/0209262",
    archivePrefix = "arXiv",
    reportNumber = "ANL-HEP-PR-02-054, EFI-02-94, FERMILAB-PUB-02-215-T",
    doi = "10.1088/1367-2630/4/1/399",
    journal = "New J. Phys.",
    volume = "4",
    pages = "99",
    year = "2002"
}

@article{Servant:2014lqa,
    author = "Servant, Geraldine",
    title = "{Status Report on Universal Extra Dimensions After LHC8}",
    eprint = "1401.4176",
    archivePrefix = "arXiv",
    primaryClass = "hep-ph",
    doi = "10.1142/S0217732315400118",
    journal = "Mod. Phys. Lett. A",
    volume = "30",
    number = "15",
    pages = "1540011",
    year = "2015"
}

@article{Kobayashi:2022xsk,
    author = "Kobayashi, Tatsuo and Otsuka, Hajime and Sakamoto, Makoto and Takeuchi, Maki and Tatsuta, Yoshiyuki and Uchida, Hikaru",
    title = "{Zero-mode wave functions by localized gauge fluxes}",
    eprint = "2211.04596",
    archivePrefix = "arXiv",
    primaryClass = "hep-th",
    reportNumber = "EPHOU-22-020, KYUSHU-HET-250, KOBE-TH-22-06",
    month = "11",
    year = "2022"
}

@article{Kaplinghat:2015aga,
    author = "Kaplinghat, Manoj and Tulin, Sean and Yu, Hai-Bo",
    title = "{Dark Matter Halos as Particle Colliders: Unified Solution to Small-Scale Structure Puzzles from Dwarfs to Clusters}",
    eprint = "1508.03339",
    archivePrefix = "arXiv",
    primaryClass = "astro-ph.CO",
    doi = "10.1103/PhysRevLett.116.041302",
    journal = "Phys. Rev. Lett.",
    volume = "116",
    number = "4",
    pages = "041302",
    year = "2016"
}

@article{Freitas:2017afm,
    author = "Freitas, Ayres and Kong, Kyoungchul and Wiegand, Daniel",
    title = "{Radiative corrections to masses and couplings in Universal Extra Dimensions}",
    eprint = "1711.07526",
    archivePrefix = "arXiv",
    primaryClass = "hep-ph",
    reportNumber = "PITT-PACC-1715",
    doi = "10.1007/JHEP03(2018)093",
    journal = "JHEP",
    volume = "03",
    pages = "093",
    year = "2018"
}

@article{Georgi:2000ks,
    author = "Georgi, Howard and Grant, Aaron K. and Hailu, Girma",
    title = "{Brane couplings from bulk loops}",
    eprint = "hep-ph/0012379",
    archivePrefix = "arXiv",
    reportNumber = "HUTP-00-A030",
    doi = "10.1016/S0370-2693(01)00408-7",
    journal = "Phys. Lett. B",
    volume = "506",
    pages = "207--214",
    year = "2001"
}

@article{Carena:2002me,
    author = "Carena, Marcela and Tait, Timothy M. P. and Wagner, C. E. M.",
    title = "{Branes and Orbifolds are Opaque}",
    eprint = "hep-ph/0207056",
    archivePrefix = "arXiv",
    reportNumber = "ANL-HEP-PR-02-043, EFI-02-93, FERMILAB-PUB-02-135-T",
    journal = "Acta Phys. Polon. B",
    volume = "33",
    pages = "2355",
    year = "2002"
}

@article{Cheng:2002iz,
    author = "Cheng, Hsin-Chia and Matchev, Konstantin T. and Schmaltz, Martin",
    title = "{Radiative corrections to Kaluza-Klein masses}",
    eprint = "hep-ph/0204342",
    archivePrefix = "arXiv",
    reportNumber = "EFI-02-73, UFIFT-HEP-02-6, BUHEP-02-20",
    doi = "10.1103/PhysRevD.66.036005",
    journal = "Phys. Rev. D",
    volume = "66",
    pages = "036005",
    year = "2002"
}

@article{Sakamura:2014aja,
    author = "Sakamura, Yutaka and Yamada, Yusuke",
    title = "{Natural realization of a large extra dimension in 5D supersymmetric theory}",
    eprint = "1401.1921",
    archivePrefix = "arXiv",
    primaryClass = "hep-ph",
    reportNumber = "KEK-TH-1699, WU-HEP-14-01",
    doi = "10.1093/ptep/ptu114",
    journal = "PTEP",
    volume = "2014",
    number = "9",
    pages = "093B02",
    year = "2014"
}

@article{Essig:2015cda,
    author = "Essig, Rouven and Fernandez-Serra, Marivi and Mardon, Jeremy and Soto, Adrian and Volansky, Tomer and Yu, Tien-Tien",
    title = "{Direct Detection of sub-GeV Dark Matter with Semiconductor Targets}",
    eprint = "1509.01598",
    archivePrefix = "arXiv",
    primaryClass = "hep-ph",
    doi = "10.1007/JHEP05(2016)046",
    journal = "JHEP",
    volume = "05",
    pages = "046",
    year = "2016"
}

@article{Castello-Mor:2020jhd,
    author = "Castell{\'o}-Mor, N.",
    editor = "Krammer, Manfred and Bergauer, Thomas and Dragicevic, Marko and Friedl, Markus and Jeitler, Manfred and Schieck, Jochen and Schwanda, Christoph",
    collaboration = "DAMIC-M",
    title = "{DAMIC-M Experiment: Thick, Silicon CCDs to search for Light Dark Matter}",
    eprint = "2001.01476",
    archivePrefix = "arXiv",
    primaryClass = "physics.ins-det",
    doi = "10.1016/j.nima.2019.162933",
    journal = "Nucl. Instrum. Meth. A",
    volume = "958",
    pages = "162933",
    year = "2020"
}

@article{Oscura:2023qik,
    author = "Cervantes-Vergara, Brenda A. and others",
    collaboration = "Oscura",
    title = "{Skipper-CCD sensors for the Oscura experiment: requirements and preliminary tests}",
    eprint = "2304.04401",
    archivePrefix = "arXiv",
    primaryClass = "physics.ins-det",
    reportNumber = "FERMILAB-PUB-23-210-PPD",
    doi = "10.1088/1748-0221/18/08/P08016",
    journal = "JINST",
    volume = "18",
    number = "08",
    pages = "P08016",
    year = "2023"
}

@inproceedings{SuperCDMS:2022kse,
    author = "Albakry, M. F. and others",
    collaboration = "SuperCDMS",
    title = "{A Strategy for Low-Mass Dark Matter Searches with Cryogenic Detectors in the SuperCDMS SNOLAB Facility}",
    booktitle = "{Snowmass 2021}",
    eprint = "2203.08463",
    archivePrefix = "arXiv",
    primaryClass = "physics.ins-det",
    reportNumber = "FERMILAB-CONF-22-171-PPD-SQMS",
    month = "3",
    year = "2022"
}

@article{Essig:2019kfe,
    author = "Essig, Rouven and P{\'e}rez-R{\'\i}os, Jes{\'u}s and Ramani, Harikrishnan and Slone, Oren",
    title = "{Direct Detection of Spin-(In)dependent Nuclear Scattering of Sub-GeV Dark Matter Using Molecular Excitations}",
    eprint = "1907.07682",
    archivePrefix = "arXiv",
    primaryClass = "hep-ph",
    doi = "10.1103/PhysRevResearch.1.033105",
    journal = "Phys. Rev. Research.",
    volume = "1",
    pages = "033105",
    year = "2019"
}

@article{Hertel:2018aal,
    author = "Hertel, S. A. and Biekert, A. and Lin, J. and Velan, V. and McKinsey, D. N.",
    title = "{Direct detection of sub-GeV dark matter using a superfluid $^4$He target}",
    eprint = "1810.06283",
    archivePrefix = "arXiv",
    primaryClass = "physics.ins-det",
    doi = "10.1103/PhysRevD.100.092007",
    journal = "Phys. Rev. D",
    volume = "100",
    number = "9",
    pages = "092007",
    year = "2019"
}

@article{Carew:2023qrj,
    author = "Carew, Ben and Caddell, Ashlee R. and Maity, Tarak Nath and O'Hare, Ciaran A. J.",
    title = "{Neutrino fog for dark matter-electron scattering experiments}",
    eprint = "2312.04303",
    archivePrefix = "arXiv",
    primaryClass = "hep-ph",
    doi = "10.1103/PhysRevD.109.083016",
    journal = "Phys. Rev. D",
    volume = "109",
    number = "8",
    pages = "083016",
    year = "2024"
}

@article{Chang:2016ntp,
    author = "Chang, Jae Hyeok and Essig, Rouven and McDermott, Samuel D.",
    title = "{Revisiting Supernova 1987A Constraints on Dark Photons}",
    eprint = "1611.03864",
    archivePrefix = "arXiv",
    primaryClass = "hep-ph",
    reportNumber = "YITP-SB-16-44",
    doi = "10.1007/JHEP01(2017)107",
    journal = "JHEP",
    volume = "01",
    pages = "107",
    year = "2017"
}

@article{Chang:2018rso,
    author = "Chang, Jae Hyeok and Essig, Rouven and McDermott, Samuel D.",
    title = "{Supernova 1987A Constraints on Sub-GeV Dark Sectors, Millicharged Particles, the QCD Axion, and an Axion-like Particle}",
    eprint = "1803.00993",
    archivePrefix = "arXiv",
    primaryClass = "hep-ph",
    reportNumber = "YITP-SB-18-01, FERMILAB-PUB-17-432-T",
    doi = "10.1007/JHEP09(2018)051",
    journal = "JHEP",
    volume = "09",
    pages = "051",
    year = "2018"
}

@article{Essig:2011nj,
    author = "Essig, Rouven and Mardon, Jeremy and Volansky, Tomer",
    title = "{Direct Detection of Sub-GeV Dark Matter}",
    eprint = "1108.5383",
    archivePrefix = "arXiv",
    primaryClass = "hep-ph",
    reportNumber = "SLAC-PUB-14538",
    doi = "10.1103/PhysRevD.85.076007",
    journal = "Phys. Rev. D",
    volume = "85",
    pages = "076007",
    year = "2012"
}

@article{DAMIC-M:2023gxo,
    author = "Arnquist, I. and others",
    collaboration = "DAMIC-M",
    title = "{First Constraints from DAMIC-M on Sub-GeV Dark-Matter Particles Interacting with Electrons}",
    eprint = "2302.02372",
    archivePrefix = "arXiv",
    primaryClass = "hep-ex",
    reportNumber = "FERMILAB-PUB-23-039-PPD",
    doi = "10.1103/PhysRevLett.130.171003",
    journal = "Phys. Rev. Lett.",
    volume = "130",
    number = "17",
    pages = "171003",
    year = "2023"
}

@article{DAMIC:2019dcn,
    author = "Aguilar-Arevalo, A. and others",
    collaboration = "DAMIC",
    title = "{Constraints on Light Dark Matter Particles Interacting with Electrons from DAMIC at SNOLAB}",
    eprint = "1907.12628",
    archivePrefix = "arXiv",
    primaryClass = "astro-ph.CO",
    reportNumber = "FERMILAB-PUB-19-432-AE-SCD",
    doi = "10.1103/PhysRevLett.123.181802",
    journal = "Phys. Rev. Lett.",
    volume = "123",
    number = "18",
    pages = "181802",
    year = "2019"
}

@article{SENSEI:2020dpa,
    author = "Barak, Liron and others",
    collaboration = "SENSEI",
    title = "{SENSEI: Direct-Detection Results on sub-GeV Dark Matter from a New Skipper-CCD}",
    eprint = "2004.11378",
    archivePrefix = "arXiv",
    primaryClass = "astro-ph.CO",
    reportNumber = "YITP-SB-2020-6, FERMILAB-PUB-20-158-AE-E",
    doi = "10.1103/PhysRevLett.125.171802",
    journal = "Phys. Rev. Lett.",
    volume = "125",
    number = "17",
    pages = "171802",
    year = "2020"
}

@article{EDELWEISS:2020fxc,
    author = "Arnaud, Q. and others",
    collaboration = "EDELWEISS",
    title = "{First germanium-based constraints on sub-MeV Dark Matter with the EDELWEISS experiment}",
    eprint = "2003.01046",
    archivePrefix = "arXiv",
    primaryClass = "astro-ph.GA",
    doi = "10.1103/PhysRevLett.125.141301",
    journal = "Phys. Rev. Lett.",
    volume = "125",
    number = "14",
    pages = "141301",
    year = "2020"
}

@article{SuperCDMS:2018mne,
    author = "Agnese, R. and others",
    collaboration = "SuperCDMS",
    title = "{First Dark Matter Constraints from a SuperCDMS Single-Charge Sensitive Detector}",
    eprint = "1804.10697",
    archivePrefix = "arXiv",
    primaryClass = "hep-ex",
    reportNumber = "FERMILAB-PUB-18-150-AE",
    doi = "10.1103/PhysRevLett.121.051301",
    journal = "Phys. Rev. Lett.",
    volume = "121",
    number = "5",
    pages = "051301",
    year = "2018",
    note = "[Erratum: Phys.Rev.Lett. 122, 069901 (2019)]"
}

@article{CDEX:2022kcd,
    author = "Zhang, Z. Y. and others",
    collaboration = "CDEX",
    title = "{Constraints on Sub-GeV Dark Matter{\textendash}Electron Scattering from the CDEX-10 Experiment}",
    eprint = "2206.04128",
    archivePrefix = "arXiv",
    primaryClass = "hep-ex",
    doi = "10.1103/PhysRevLett.129.221301",
    journal = "Phys. Rev. Lett.",
    volume = "129",
    number = "22",
    pages = "221301",
    year = "2022"
}

@article{DarkSide:2022knj,
    author = "Agnes, P. and others",
    collaboration = "DarkSide",
    title = "{Search for Dark Matter Particle Interactions with Electron Final States with DarkSide-50}",
    eprint = "2207.11968",
    archivePrefix = "arXiv",
    primaryClass = "hep-ex",
    reportNumber = "FERMILAB-PUB-22-566-ND-PPD-SCD",
    doi = "10.1103/PhysRevLett.130.101002",
    journal = "Phys. Rev. Lett.",
    volume = "130",
    number = "10",
    pages = "101002",
    year = "2023"
}

@article{XENON:2019gfn,
    author = "Aprile, E. and others",
    collaboration = "XENON",
    title = "{Light Dark Matter Search with Ionization Signals in XENON1T}",
    eprint = "1907.11485",
    archivePrefix = "arXiv",
    primaryClass = "hep-ex",
    doi = "10.1103/PhysRevLett.123.251801",
    journal = "Phys. Rev. Lett.",
    volume = "123",
    number = "25",
    pages = "251801",
    year = "2019"
}

@article{XENON:2021qze,
    author = "Aprile, E. and others",
    collaboration = "XENON",
    title = "{Emission of single and few electrons in XENON1T and limits on light dark matter}",
    eprint = "2112.12116",
    archivePrefix = "arXiv",
    primaryClass = "hep-ex",
    doi = "10.1103/PhysRevD.106.022001",
    journal = "Phys. Rev. D",
    volume = "106",
    number = "2",
    pages = "022001",
    year = "2022",
    note = "[Erratum: Phys.Rev.D 110, 109903 (2024)]"
}

@article{PandaX-II:2021nsg,
    author = "Cheng, Chen and others",
    collaboration = "PandaX-II",
    title = "{Search for Light Dark Matter-Electron Scatterings in the PandaX-II Experiment}",
    eprint = "2101.07479",
    archivePrefix = "arXiv",
    primaryClass = "hep-ex",
    doi = "10.1103/PhysRevLett.126.211803",
    journal = "Phys. Rev. Lett.",
    volume = "126",
    number = "21",
    pages = "211803",
    year = "2021"
}

@article{Belle-II:2018jsg,
    author = "Altmannshofer, W. and others",
    editor = "Kou, E. and Urquijo, P.",
    collaboration = "Belle-II",
    title = "{The Belle II Physics Book}",
    eprint = "1808.10567",
    archivePrefix = "arXiv",
    primaryClass = "hep-ex",
    reportNumber = "KEK Preprint 2018-27, BELLE2-PUB-PH-2018-001, FERMILAB-PUB-18-398-T, JLAB-THY-18-2780, INT-PUB-18-047, UWThPh 2018-26",
    doi = "10.1093/ptep/ptz106",
    journal = "PTEP",
    volume = "2019",
    number = "12",
    pages = "123C01",
    year = "2019",
    note = "[Erratum: PTEP 2020, 029201 (2020)]"
}

@article{Gninenko:2019qiv,
    author = "Gninenko, S. N. and Kirpichnikov, D. V. and Kirsanov, M. M. and Krasnikov, N. V.",
    title = "{Combined search for light dark matter with electron and muon beams at NA64}",
    eprint = "1903.07899",
    archivePrefix = "arXiv",
    primaryClass = "hep-ph",
    doi = "10.1016/j.physletb.2019.07.015",
    journal = "Phys. Lett. B",
    volume = "796",
    pages = "117--122",
    year = "2019"
}

@article{Anchordoqui:2021ghd,
    author = "Anchordoqui, Luis A. and others",
    title = "{The Forward Physics Facility: Sites, experiments, and physics potential}",
    eprint = "2109.10905",
    archivePrefix = "arXiv",
    primaryClass = "hep-ph",
    reportNumber = "BNL-222142-2021-FORE, CERN-PBC-Notes-2021-025, DESY-21-142, DESY-21-142,
  FERMILAB-CONF-21-452-AE-E-ND-PPD-T, KYUSHU-RCAPP-2021-01, LU TP 21-36,
  PITT-PACC-2118, SMU-HEP-21-10, UCI-TR-2021-22, FERMILAB-CONF-21-452-AE-E-ND-PPD-T",
    doi = "10.1016/j.physrep.2022.04.004",
    journal = "Phys. Rept.",
    volume = "968",
    pages = "1--50",
    year = "2022"
}

@article{Battaglieri:2020lds,
    author = "Battaglieri, M. and others",
    title = "{The BDX-MINI detector for Light Dark Matter search at JLab}",
    eprint = "2011.10532",
    archivePrefix = "arXiv",
    primaryClass = "physics.ins-det",
    doi = "10.1140/epjc/s10052-021-08957-5",
    journal = "Eur. Phys. J. C",
    volume = "81",
    number = "2",
    pages = "164",
    year = "2021"
}

@inproceedings{Akesson:2022vza,
    author = "{\r{A}}kesson, Torsten and others",
    title = "{Current Status and Future Prospects for the Light Dark Matter eXperiment}",
    booktitle = "{Snowmass 2021}",
    eprint = "2203.08192",
    archivePrefix = "arXiv",
    primaryClass = "hep-ex",
    reportNumber = "FERMILAB-CONF-22-313-PPD-SCD-T",
    month = "3",
    year = "2022"
}

@article{LDMX:2018cma,
    author = "{\r{A}}kesson, Torsten and others",
    collaboration = "LDMX",
    title = "{Light Dark Matter eXperiment (LDMX)}",
    eprint = "1808.05219",
    archivePrefix = "arXiv",
    primaryClass = "hep-ex",
    reportNumber = "FERMILAB-PUB-18-324-A, SLAC-PUB-17303",
    month = "8",
    year = "2018"
}

@article{BaBar:2017tiz,
    author = "Lees, J. P. and others",
    collaboration = "BaBar",
    title = "{Search for Invisible Decays of a Dark Photon Produced in ${e}^{+}{e}^{-}$ Collisions at BaBar}",
    eprint = "1702.03327",
    archivePrefix = "arXiv",
    primaryClass = "hep-ex",
    reportNumber = "BABAR-PUB-17-001, SLAC-PUB-16923",
    doi = "10.1103/PhysRevLett.119.131804",
    journal = "Phys. Rev. Lett.",
    volume = "119",
    number = "13",
    pages = "131804",
    year = "2017"
}

@article{CMS:2021far,
    author = "Tumasyan, Armen and others",
    collaboration = "CMS",
    title = "{Search for new particles in events with energetic jets and large missing transverse momentum in proton-proton collisions at $ \sqrt{s} $ = 13 TeV}",
    eprint = "2107.13021",
    archivePrefix = "arXiv",
    primaryClass = "hep-ex",
    reportNumber = "CMS-EXO-20-004, CERN-EP-2021-136",
    doi = "10.1007/JHEP11(2021)153",
    journal = "JHEP",
    volume = "11",
    pages = "153",
    year = "2021"
}

@article{Andreev:2021fzd,
    author = "Andreev, Yu. M. and others",
    title = "{Improved exclusion limit for light dark matter from e+e- annihilation in NA64}",
    eprint = "2108.04195",
    archivePrefix = "arXiv",
    primaryClass = "hep-ex",
    reportNumber = "CERN-EP-2021-164",
    doi = "10.1103/PhysRevD.104.L091701",
    journal = "Phys. Rev. D",
    volume = "104",
    number = "9",
    pages = "L091701",
    year = "2021"
}

@article{Batell:2014mga,
    author = "Batell, Brian and Essig, Rouven and Surujon, Ze'ev",
    title = "{Strong Constraints on Sub-GeV Dark Sectors from SLAC Beam Dump E137}",
    eprint = "1406.2698",
    archivePrefix = "arXiv",
    primaryClass = "hep-ph",
    doi = "10.1103/PhysRevLett.113.171802",
    journal = "Phys. Rev. Lett.",
    volume = "113",
    number = "17",
    pages = "171802",
    year = "2014"
}

@article{deNiverville:2011it,
    author = "deNiverville, Patrick and Pospelov, Maxim and Ritz, Adam",
    title = "{Observing a light dark matter beam with neutrino experiments}",
    eprint = "1107.4580",
    archivePrefix = "arXiv",
    primaryClass = "hep-ph",
    doi = "10.1103/PhysRevD.84.075020",
    journal = "Phys. Rev. D",
    volume = "84",
    pages = "075020",
    year = "2011"
}

@article{MiniBooNEDM:2018cxm,
    author = "Aguilar-Arevalo, A. A. and others",
    collaboration = "MiniBooNE DM",
    title = "{Dark Matter Search in Nucleon, Pion, and Electron Channels from a Proton Beam Dump with MiniBooNE}",
    eprint = "1807.06137",
    archivePrefix = "arXiv",
    primaryClass = "hep-ex",
    reportNumber = "LA-UR-18-26421, FERMILAB-PUB-18-334-ND",
    doi = "10.1103/PhysRevD.98.112004",
    journal = "Phys. Rev. D",
    volume = "98",
    number = "11",
    pages = "112004",
    year = "2018"
}

@article{Barbieri:2000vh,
    author = "Barbieri, Riccardo and Hall, Lawrence J. and Nomura, Yasunori",
    title = "{A Constrained standard model from a compact extra dimension}",
    eprint = "hep-ph/0011311",
    archivePrefix = "arXiv",
    reportNumber = "SNS-PH-00-16, UCB-PTH-00-38, LBNL-47128",
    doi = "10.1103/PhysRevD.63.105007",
    journal = "Phys. Rev. D",
    volume = "63",
    pages = "105007",
    year = "2001"
}
\bibliographystyle{utcaps_mod}
\clearpage
\onecolumngrid

\begin{center}
  \large\bfseries
  Supplemental Material\\
  Naturally Resonant Dark Matter from Extra Dimensions
\end{center}

\setcounter{equation}{0}
\renewcommand{\theequation}{S\arabic{equation}}
\setcounter{figure}{0}
\renewcommand{\thefigure}{S\arabic{figure}}
\setcounter{table}{0}
\renewcommand{\thetable}{S\Roman{table}}
\setcounter{section}{0}
\setcounter{subsection}{0}
\renewcommand{\thesection}{S\Roman{section}}

\ifdefined\theHequation
  \renewcommand{\theHequation}{supp.\arabic{equation}}
\fi
\ifdefined\theHfigure
  \renewcommand{\theHfigure}{supp.\arabic{figure}}
\fi
\ifdefined\theHtable
  \renewcommand{\theHtable}{supp.\arabic{table}}
\fi
\ifdefined\theHsection
  \renewcommand{\theHsection}{supp.\arabic{section}}
\fi

\section{Details of the Level of Resonance}
\label{supp:app:a}

\begin{figure*}[ht]
\centering
\includegraphics[scale=0.282]{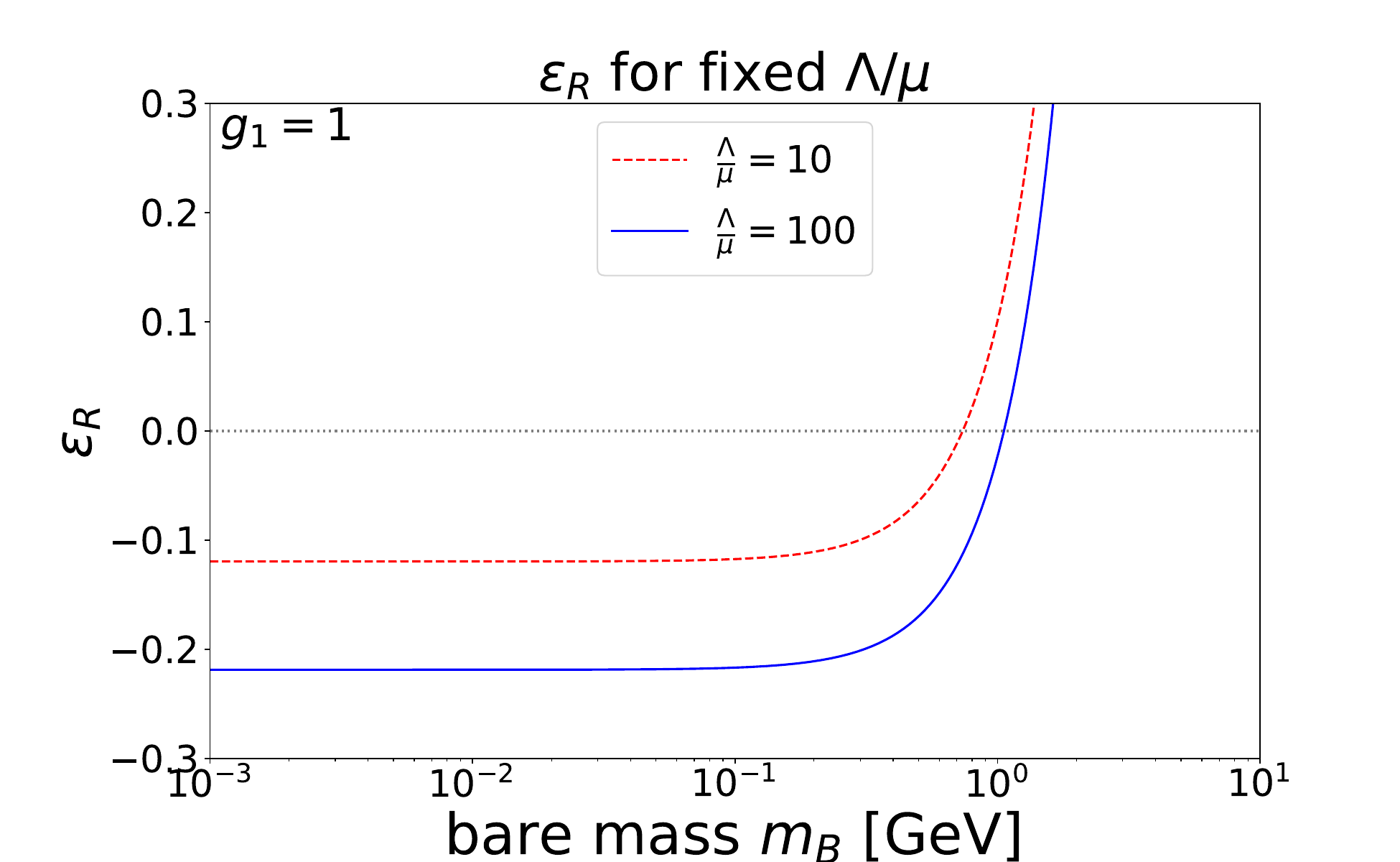}\hspace{-2.3em}
\includegraphics[scale=0.282]{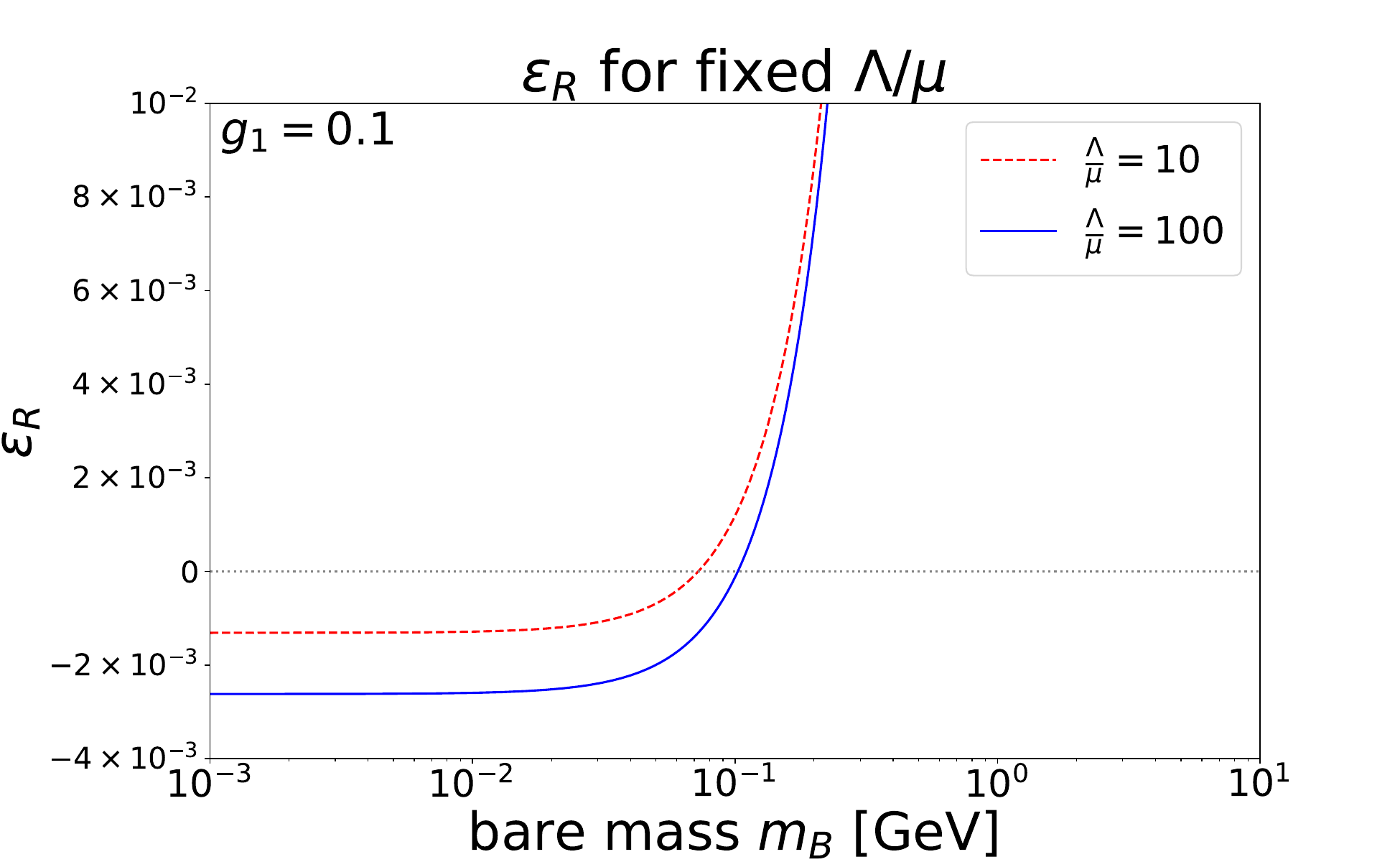}\hspace{-2.3em}
\includegraphics[scale=0.282]
{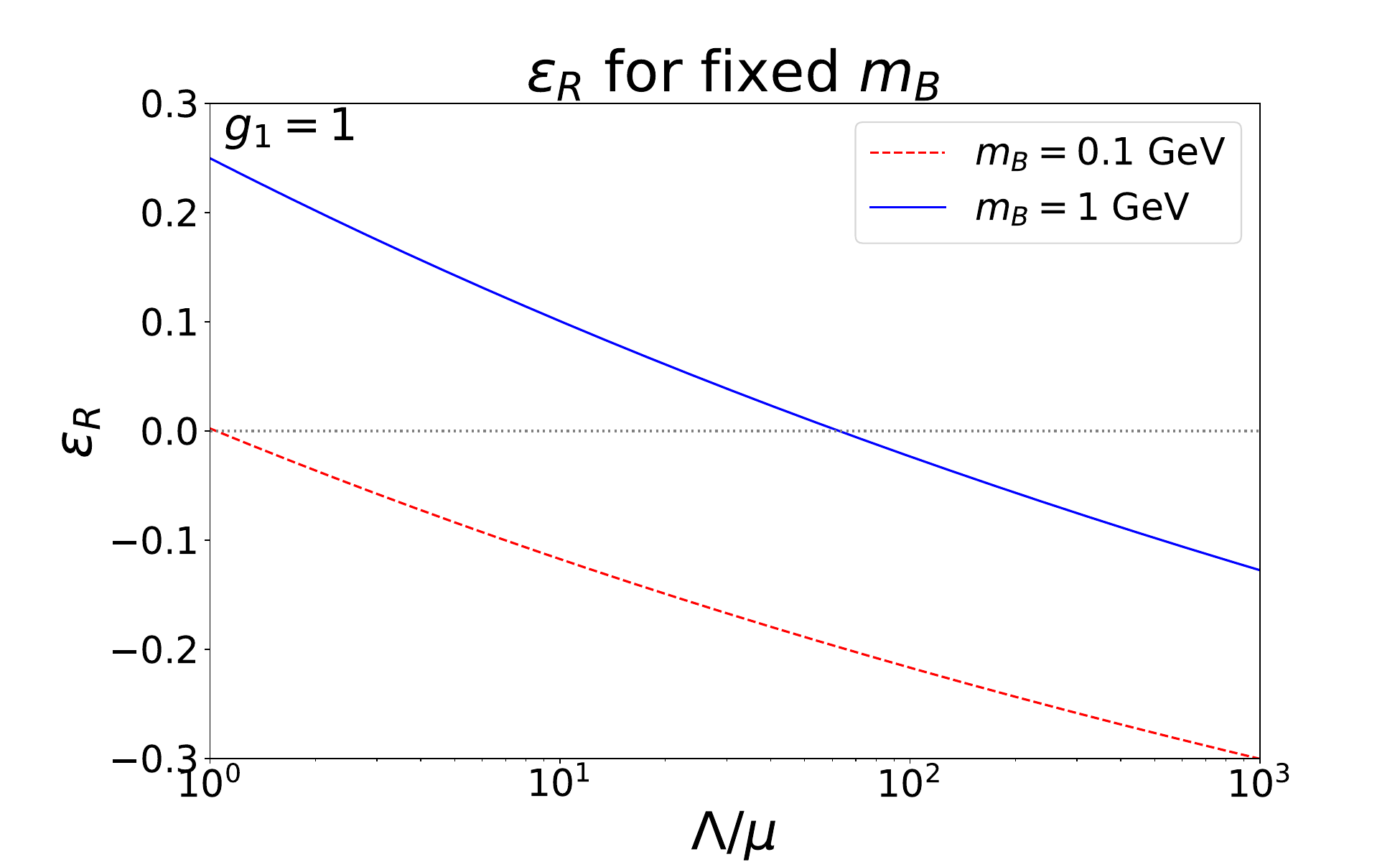}\hspace{-2.3em}
\includegraphics[scale=0.282]{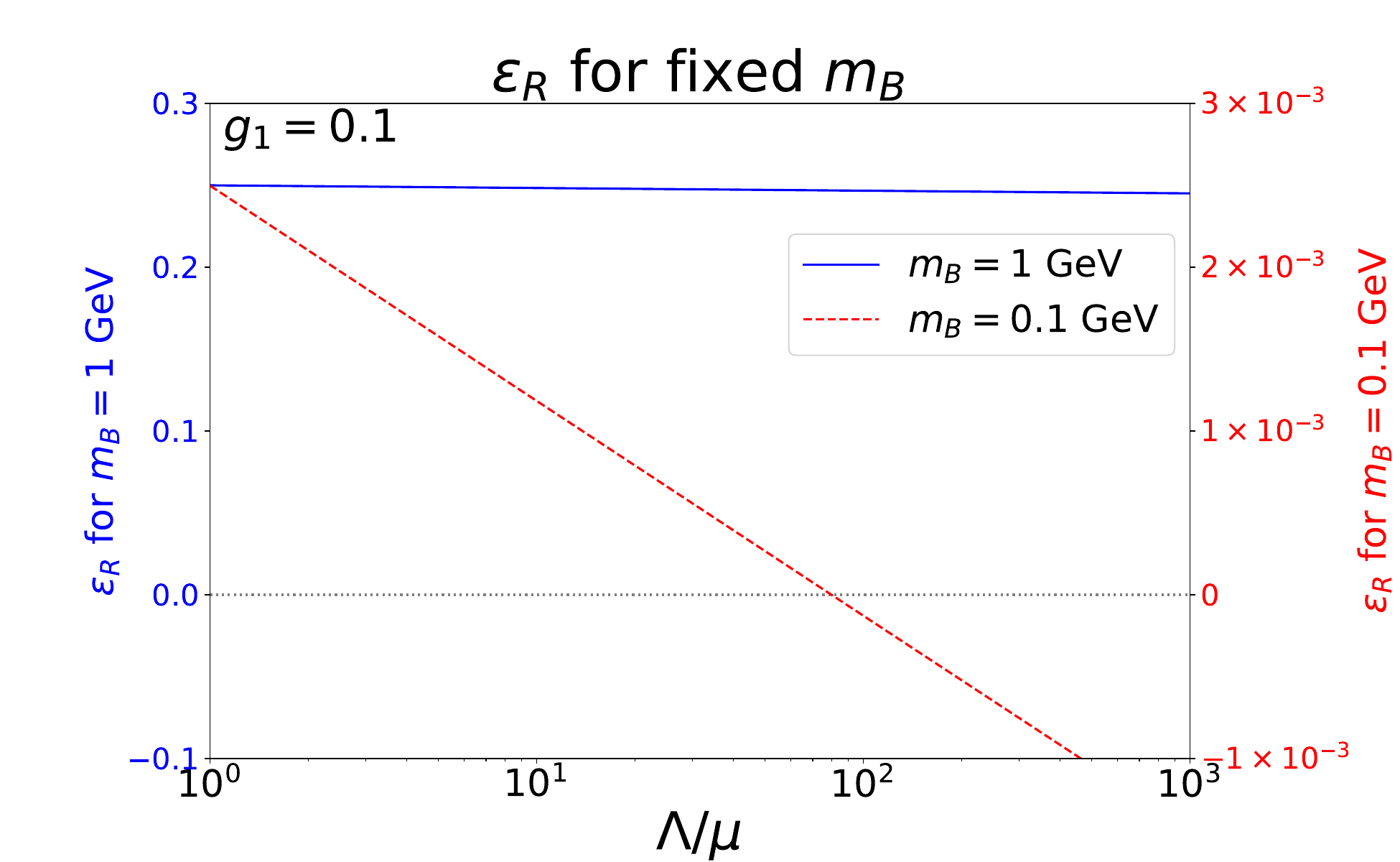}
\caption{
The two panels on the top show the level of resonance, $\epsilon_R$ (defined in the text), in terms of $m_B$ for fixed $\Lambda/{\mu}$ values, for $g_1 =1$ (left), and $g_1 =0.1$  (right). The two panels on the bottom show the level of resonance in terms of $\Lambda/{\mu}$ for fixed $m_B$ values, for $g_1 =1$ (left), and $g_1 =0.1$  (right). Note that for the lower right panel, two different units have been chosen for $\epsilon_R$ for $m_B=1\;{\rm GeV}$ (left vertical axis) and $m_B=0.1\;{\rm GeV}$ (right vertical axis) due to significant difference in the values of $\epsilon_R$. The compactification scale, $1/R$, is fixed to ${1}/{R}=1\;{\rm GeV}$ when making all 4 panels. 
}
\label{supp:fermion_levelofresonance_S1Z2Z2g1}
\end{figure*}
In Fig.~\ref{supp:fermion_levelofresonance_S1Z2Z2g1}, we present the behavior of the resonance level $\epsilon_R$ as a function of the bare mass $m_B$, keeping $\Lambda / \mu$ fixed. The other model parameters are set to $g_1 = 0.1, 1$ and the compactification scale is chosen as $1 / R = 1~\text{GeV}$. For fixed $\Lambda / \mu$, the level of resonance is nearly constant for bare mass values $m_B \leq 1 / R$, with $\epsilon_R \sim -\mathcal{O}(0.1)$. However, achieving $\epsilon_R > 0$ with a magnitude of $\mathcal{O}(0.1)$ is more challenging and may require fine-tuning to achieve the resonance phenomenology~\cite{Feng:2017drg} with $g_1 = 1$. Notably, for the smaller coupling $g_1 = 0.1$, and in the region where $m_B \sim 1 / R$, $\epsilon_R$ remains less than or comparable to $\mathcal{O}(0.1)$ over a broad range of values for the cutoff ratio $\Lambda / \mu$. However, to obtain much larger resonance, $\epsilon_R < \mathcal{O}(0.1)$, the bare mass must lie in the range $0.07\times(1/R) < m_B < 0.7\times(1/R)$. This characteristic allows the model to exhibit a resonance across a large region of parameter space, while consistently maintaining the invisibility of the dark photon.

\begin{figure*}[ht]
\centering
\includegraphics[scale=0.282]{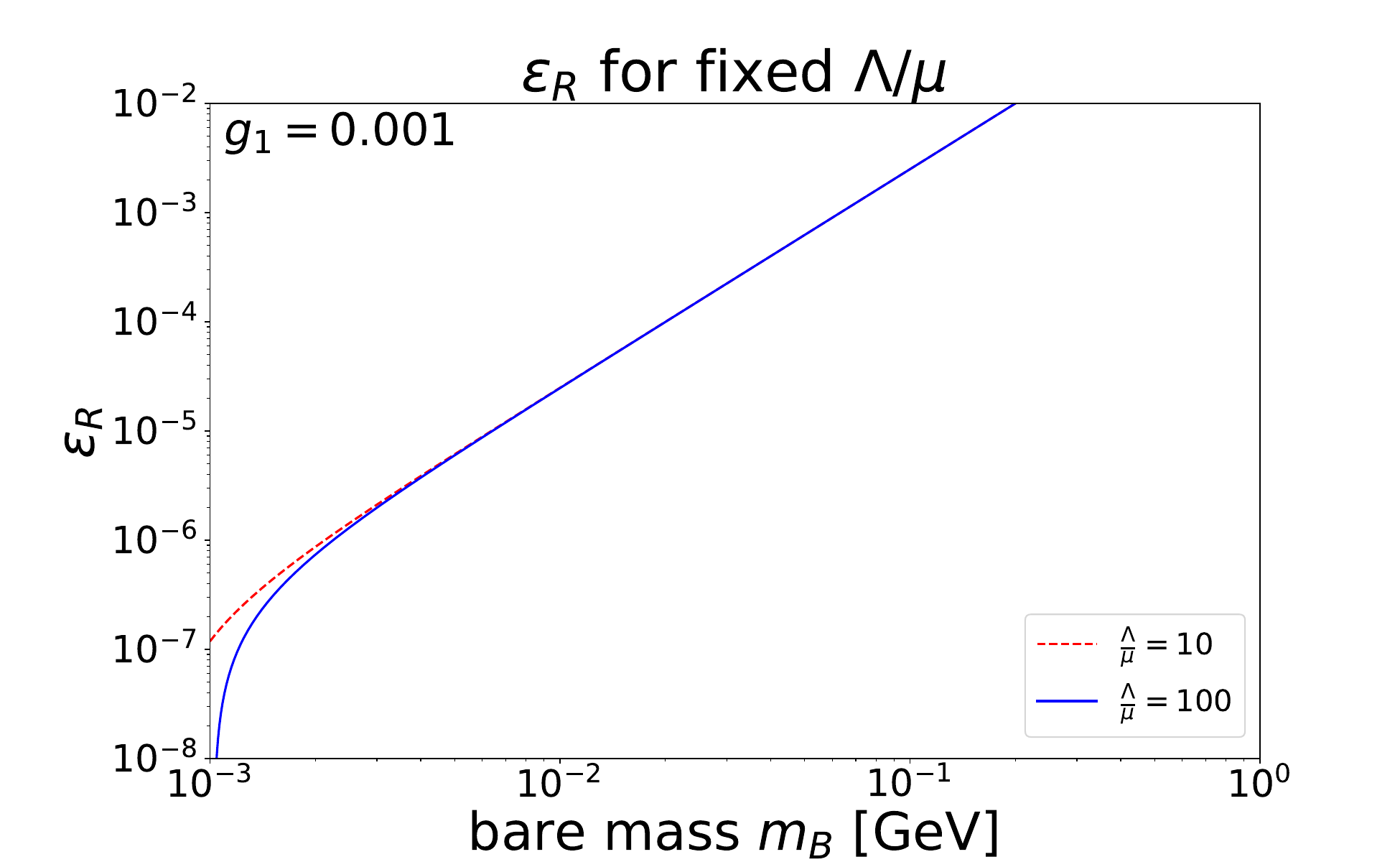}\hspace{-2.3em}
\includegraphics[scale=0.282]
{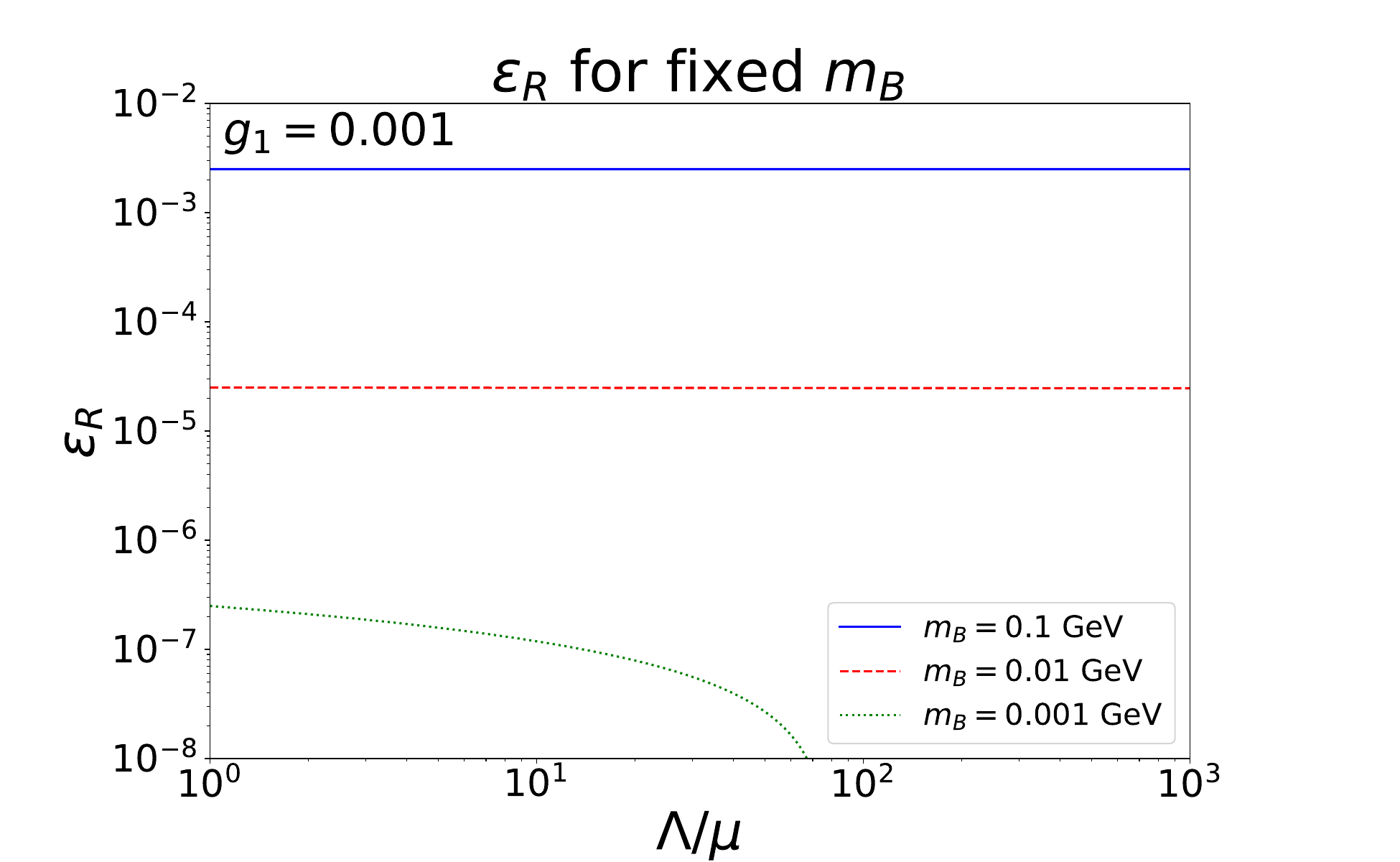}\hspace{-2.3em}
\caption{
The left panel shows the resonance level, $\epsilon_R$,
as a function of $m_B$ for fixed $\Lambda/\mu$.
The right panel shows $\epsilon_R$ as a function of
$\Lambda/\mu$ for fixed $m_B$.
Both panels use $g_1=0.001$ and $1/R=1~\mathrm{GeV}$.
}
\label{supp:fermion_levelofresonance_for_small_g1}
\end{figure*}

In Fig.~\ref{supp:fermion_levelofresonance_for_small_g1}, we focus on the resonance regime $10^{-8}\lesssim \epsilon_R \lesssim 10^{-4}$, which can alleviate small-scale structure problems via self-interactions~\cite{Tsai:2020vpi,Chu:2018fzy}. Achieving this regime requires a very small coupling, so we fix $g_1=0.001$ and illustrate how minimally the remaining parameters must be tuned to realize such a resonance. We also explain in the main text why taking such a small $g_1$ does not necessarily mean fine-tuning.

\section{Details of the $S^1/(Z_2\times Z_2')$ Orbifold}
\label{supp:App:A}

\begin{figure*}[t]
\includegraphics[scale=0.4]{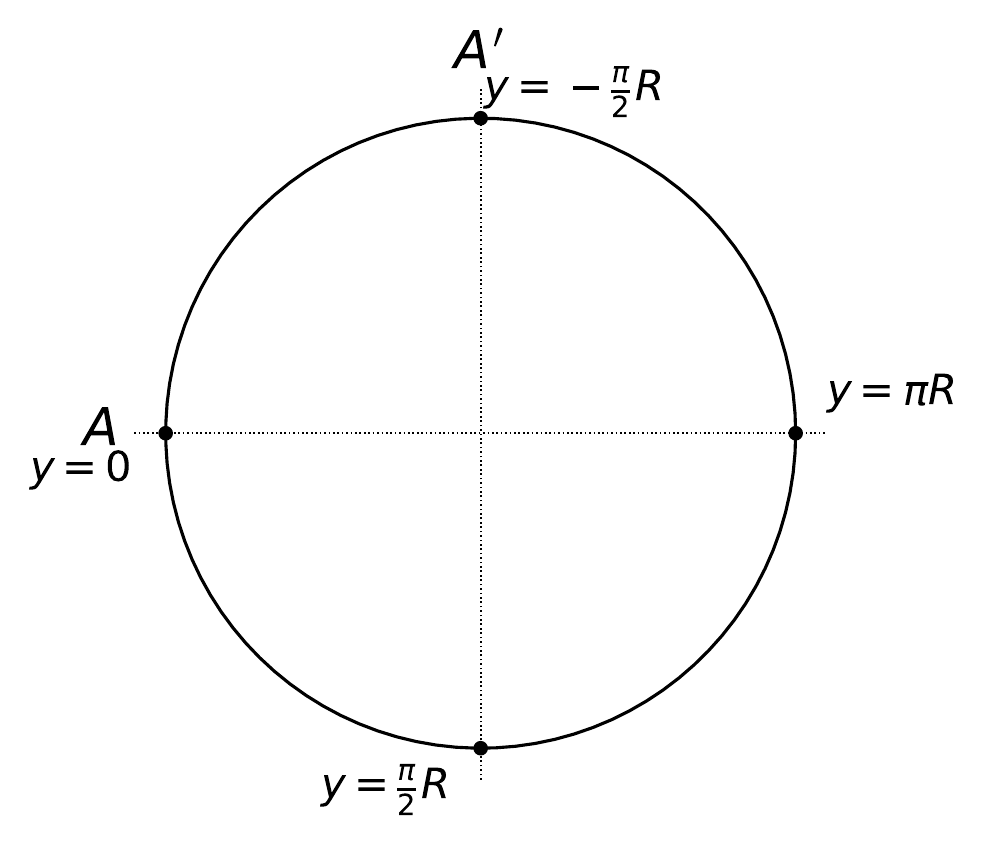}
\caption{$S^1/(Z_2\times Z_2')$ orbifold in fifth dimension \cite{Barbieri:2000vh}.}
\label{supp:s1z2z2orb}
\end{figure*}

The $S^1/(Z_2\times Z_2')$ has two parities imposed on a circle~\cite{Barbieri:2000vh} (see Fig.~\ref{supp:s1z2z2orb}). In addition to parity, $Z_2 : y\rightarrow -y $ (reflection about the axes $A$), $Z_2' : y'\rightarrow -y' $ (reflection about the axes $A'$) is imposed, where $y'=y-\pi R/2$. Four modes are assembled on a circle with a specific $(Z_2,Z_2')$ quantum number. Specific boundary conditions explained in section {\bf Fermionic Dark Matter Model} of the main text can also be interpreted in terms of the quantum number for the discrete symmetry. The Dirichlet boundary condition corresponds to $(-)$ quantum number, whereas the Neumann boundary condition corresponds to $(+)$ quantum number for the $Z_2(Z_2')$ symmetry. Table I of the main text can be rewritten in terms of the quantum number given in Table~\ref{supp:table:BCS1Z2Z2_QN}, which matches with the discussions in~\cite{Barbieri:2000vh}.

\begin{table}[h!]
\centering
\begin{tabular}{|c | c | c | } 
 \hline
 symmetry &     $\;\;\;\;\;\;Z_2\;\;\;\;\;\;$     &     $\;\;\;\;\;\;Z_2'\;\;\;\;\;\;$     \\ [1ex]
 \hline
RH Singlet fermion & + & $-$ \\
 \hline
 LH Singlet fermion & $-$ & + \\
 \hline
 Dark Photon ($B_\mu$) & + & + \\ 
\hline
Dark Photon ($B_5$) & $-$ & $-$ \\ 
\hline
\end{tabular}
\caption{Summary of the quantum numbers for KK particles.
}
\label{supp:table:BCS1Z2Z2_QN}
\end{table}

\section{Details of relic abundance calculation}
The non-relativistic cross section can be approximated to leading order in velocity~\cite{Feng:2017drg} as
\begin{eqnarray}
    \langle\sigma v\rangle_{\rm NR} &\simeq& \frac{2x^{3/2}\pi^{1/2}}{s_0}F^{(1)} \\
     &&\cdot \left(\gamma_R\pi^{-1/2}x^{-1/2} +{\rm Re}\left[z_R^{3/2}w\left(x^{1/2}z_R^{1/2}\right)\right] \right), \nonumber
\end{eqnarray}
where variables are defined as
\begin{eqnarray}
    s_0 &=& 4m_{E^{(1)}}'^2, \\
    x &=& {m_{E^{(1)}}'}/{T}, \\
    \gamma_R &=& m'_{B^{(2)}}\Gamma_{B^{(2)}}/s_0, \\
    z_R &=& \epsilon_R + i \gamma_R,
\end{eqnarray}
and $w(z)$ is the Faddeeva function.

The form factor $F^{(1)}$ is calculated to be
\begin{equation}
    F^{(1)} =  \frac{8\pi}{3} \frac{\kappa^2 \alpha_E\alpha_D}{\gamma_R} \frac{1}{B_e(2m_{E^{(1)}}')},
\end{equation}
where $\alpha_E$ is electromagnetic fine structure constant and $B_e$ is defined to be 
\begin{equation}
    B_e = \frac{\Gamma(B^{(2)}\rightarrow ee)}{\Gamma(B^{(2)} \rightarrow {\rm SM})}.
\end{equation}
Assuming that dark photons predominantly decay into dark matter (DM), $\gamma_R$ can be approximated as
\begin{equation}
    \gamma_R \simeq \frac{\alpha_D}{3}\sqrt{\frac{1}{1+\epsilon_R}}\epsilon_R^{3/2}.
\end{equation}
The thermal relic abundance of DM is given by
\begin{equation}
    \Omega_Dh^2 \approx 8.77\times 10^{-11}\;{\rm GeV^{-2}}\left[\bar{g}_{\rm eff}\int_{x_f}^{x_0}\frac{\langle\sigma v\rangle_{\rm NR}}{x^2}dx\right]^{-1},
\end{equation}
where $x_0$ and $x_f$ are calculated from the current temperature of the universe and the temperature at freeze-out, respectively. The freeze-out condition determines the freeze-out temperature,
\begin{equation}
    m_Dm_{\rm Pl}\langle\sigma v\rangle_{\rm NR}
     \approx 7.043\times x_f^{1/2}e^{x_f}\frac{h_{\rm eff}(x_f)}{g_{\rm eff}^{1/2}(x_f)},
\end{equation}
where $g_{\rm eff}$ and $h_{\rm eff}$ represent the effective degrees of freedom for energy and entropy, respectively. Finally, $\bar{g}_{\rm eff}$ is the typical value of the effective degrees of freedom for energy between freeze-out and the present universe.

\end{document}